\documentclass[aps,preprint,superscriptaddress,]{revtex4}
\usepackage{graphics}
\usepackage{graphicx}
\begin{document}
\title{A nonplanar Peierls-Nabarro model and its applications to dislocation
 cross-slip} 
\author{Gang Lu}
\affiliation
{Department of Physics and Division of Engineering and Applied Science,
 Harvard University, Cambridge, MA 02138}
\author{Vasily V. Bulatov}
\affiliation{Lawrence Livermore National Laboratory, Livermore, CA 94550}
\author{Nicholas Kioussis}
\affiliation{Department of Physics, California State University Northridge,
Northridge, CA 91330}
\begin{abstract} 
\vskip 0.5cm
A novel semidiscrete Peierls-Nabarro model is introduced which can be used to
study dislocation spreading at more than one slip planes, such as dislocation
cross-slip and junctions. The strength of the model, when combined with
{\it ab initio} calculations for the energetics, is that it produces essentially
an atomistic simulation for dislocation core properties without
suffering from the uncertainties associated with empirical potentials.
Therefore, this method is particularly useful in providing insight into
alloy design when empirical
potentials are not available or not reliable for such multi-element systems.   
As an example, we study dislocation cross-slip and
constriction process in two contrasting fcc metals, Al and Ag.
We find that the screw dislocation in Al can cross-slip 
spontaneously in contrast with that in Ag, where the screw dislocation
splits into two partials, which cannot cross-slip without first being constricted.
The response of the dislocation to an external stress is examined in detail.
The dislocation constriction energy and the critical stress for 
cross-slip are determined, and from the latter, we estimate the cross-slip 
energy barrier for 
straight screw dislocations. 
\end{abstract}
\maketitle
The past decades have witnessed a growing interest
towards a quantitative understanding of  
deformation and strength of materials 
and, in particular, the effect of impurities and alloying 
elements on dislocation properties.
While continuum elasticity theory describes well the
long-range elastic strain of a dislocation for length scales beyond
a few lattice spacings, it breaks down near
the singularity in the region surrounding the dislocation center,
known as the dislocation core. There has been a great deal of interest in
 describing accurately
the dislocation core structure on an
atomic scale because of its important role in many phenomena
of crystal plasticity \cite{Duesbery1,Vitek}.
The core properties control, for instance, the mobility of
dislocations, which
accounts for the intrinsic ductility or brittleness of solids.
The core is also responsible for the interaction of
dislocations at close distances, which is relevant to plastic deformation.
For example, by integrating the local rules derived
from atomistic simulations of core interactions
into dislocation-dynamics simulations, a
connection between micro-to-meso scales can be
established to study dislocation reactions
and crystal plasticity \cite{Nature}.

Cross-slip, the process by which a screw dislocation moves from one 
glide plane to another, is ubiquitous in plastic deformation
of materials. For example, cross-slip is considered to be responsible for 
the onset of stage III in the stress-strain work-hardening curves.
Furthermore, cross-slip can result in the formation of glide plane obstacles 
(sessile segments) in hcp metals and in L1$_2$-, B2- and L1$_0$- based
intermetallic alloys, responsible for the anomalous high-temperature 
yield stress peak.  
However, theoretical studies of alloying and impurity effects on
dislocation cross-slip have proved to be particularly difficult 
because one has to deal with both long-ranged elastic 
interactions (between dislocation segments) and short-ranged atomic 
interactions (due to the constriction process) that are inherent
in a cross-slip process.  

There are currently two theoretical
approaches to study cross-slip. One is based on
the line tension approximation which completely ignores atomic 
interactions, therefore it is not reliable \cite{Friedel,Escaig}. 
The other approach is direct 
atomistic simulations employing empirical potentials \cite{Rasmussen,Rao}.
Although the second approach is quite powerful in determining cross-slip 
transition paths and estimating the corresponding activation energy barriers, 
it is time-consuming and, more importantly, 
it critically depends on the accuracy and availability
of the empirical potentials employed in the simulations. For example,
reliable interatomic potentials usually are not available for multi-elements
materials. As a consequence, the possible hardening mechanisms due to
alloying for most materials are still uncertain, and the design of new 
materials based on favorable cross-slip properties lacks guidance.
Thus, the understanding at an atomic level of the chemistry effect
on the dislocation core properties and cross-slip mechanisms 
is of great importance in predicting and controlling plastic deformation 
in structural materials, since the deformation behavior  
is often associated with the presence of substitutional or interstitial 
alloying elements. The ultimate goal of theoretical studies is then to 
use this information to direct alloying design for new materials with desired 
mechanical properties by tailoring the dislocation 
properties - in close collaboration with experimental efforts.

In this paper, we introduce a novel model based on 
the Peierls-Nabarro (P-N) framework 
which allows the study of dislocation cross-slip employing 
{\it ab initio} calculations.
In fact, there has been a resurgence of interest recently in 
applying the simple and tractable P-N model to study 
dislocation core structure and mobility in conjunction with {\it ab initio} 
$\gamma$-surface calculations
 \cite{Joos,Juan,Bulatov,Hartford,Lu1,Lu3}. This approach represents a
combination of an atomistic ({\it ab initio}) treatment of the interactions 
across the slip plane and an elastic treatment of the continua 
on either side of the slip plane.  
Therefore, this approach is particularly
useful for studying the interaction of impurities with dislocations when
empirical potentials are either not available or not reliable to deal with such
multi-element systems. Furthermore, it allows to study general trends in 
dislocation core properties and to correlate them with specific features 
of the underlying electronic structure.
However, to date, all models based on the P-N framework are 
applicable only to a single slip plane while 
the important cross-slip process
requires at least two active intersecting slip planes, i.e., the primary and 
cross-slip planes. In this work the semidiscrete 
variational P-N model\cite{Bulatov,Lu1,Lu3} 
is extended so as to take into account two intersecting slip planes.
We shall apply this new model 
to study the dislocation constriction and cross-slip process 
in two fcc metals, Al and Ag, exhibiting different deformation 
properties. We are particularly interested in the evolution of 
the dislocation core structure under external stress and the 
interplay between the applied stress and the cross-slip process.

We begin by developing an appropriate energy functional for 
the Peierls dislocation at two intersecting slip planes. 
To facilitate presentation, we adopt the following conventions: 
In Fig. 1, a screw dislocation placed at the intersection of the 
primary (plane I) and cross-slip plane (plane II) is allowed to spread into the
two planes simultaneously. The $X$ ($X'$) axis represents the glide
direction of the dislocation at the plane I (II). For an fcc lattice, 
the two slip planes are (111) and ($\bar{1}11$), forming an
angle $\theta \approx$ 71$^\circ$. 
The dislocation line is along the [10$\bar{1}$] ($Z$ axis) direction and $L$ 
represents the outer radius of the dislocation beyond which the 
configuration independent elastic energy is ignored.\cite{Lu1}
 In the spirit of the P-N model, the dislocation is represented
as a continuous distribution of infinitesimal dislocations with densities of
$\rho^{\rm I}(x)$ and $\rho^{\rm II}(x')$ on the primary and 
cross-slip planes, respectively, 
where, $x$ and $x'$ are the coordinates of the atomic rows at
the two planes. Following the semidiscrete Peierls framework
developed earlier \cite{Bulatov,Lu1}, the total energy
of the dislocation is 
\begin{equation}
U_{tot}=U_{\rm I}+U_{\rm II}+\tilde{U}. 
\end{equation}
Here, $U_{\rm I}$ and $U_{\rm II}$
are the energies associated with the dislocation spread on planes I and
II, respectively, and $\tilde{U}$ represents the elastic interaction energy
between the dislocation densities on planes I and II. The expressions 
for $U_{\rm I}$ and 
$U_{\rm II}$ are identical to that given earlier for the single glide
plane case \cite{Bulatov, Lu1}, while the new term $\tilde{U}$ 
can be derived from Nabarro's equation for general parallel 
dislocations \cite{Nabarro},    
\begin{eqnarray*}
U_{\rm I(II)}&=&\sum\limits_{i,j}\frac{1}{2}\chi_{ij}\{K_e[\rho^{\rm I(II)}_1(i)
\rho^{\rm I(II)}_1(j)+\rho^{\rm I(II)}_2(i)
\rho^{\rm I(II)}_2(j)]+
K_s\rho^{\rm I(II)}_3(i)\rho^{\rm I(II)}_3(j)\}\\
& & +\sum\limits_{i} \Delta x\gamma_3\left(f^{\rm I(II)}_1(i),f^{\rm I(II)}_2(i),
f^{\rm I(II)}_3(i)\right)
-\sum\limits_{i,l}\frac{x(i)^2-x(i-1)^2}{2}
\rho^{\rm I(II)}_l(i)\tau^{\rm I(II)}_l+Kb^2{\rm ln}L,\\
\tilde{U}&=&-\sum\limits_{i,j}K_s\rho^{\rm I}_3(i)\rho^{\rm p}_3(j)
A_{ij}-\sum\limits_{i,j}
K_e[\rho^{\rm I}_1(i)\rho^{\rm p}_1(j)+\rho^{\rm I}_2(i)\rho^{\rm p}_2(j)]A_{ij}\\
& &-\sum\limits_{i,j}
K_e[\rho^{\rm I}_2(i)\rho^{\rm p}_2(j)B_{ij}+\rho^{\rm I}_1(i)
\rho^{\rm p}_1(j)C_{ij}-
\rho^{\rm I}_2(i)\rho^{\rm p}_1(j)D_{ij}-\rho^{\rm I}_1(i)
\rho^{\rm p}_2(j)D_{ij}]~.
\end{eqnarray*}
Here, $f^{\rm I(II)}_1(i)$, $f^{\rm I(II)}_2(i)$ and 
$f^{\rm I(II)}_3(i)$ represent the edge, vertical and screw component
of the general dislocation displacement at the $i$-th nodal 
point in plane I(II), respectively, while the corresponding 
component of dislocation density in plane I(II) is defined as 
$\rho^{\rm I(II)}(i) = \left(f^{\rm I(II)}(i)-
f^{\rm I(II)}(i-1)\right)/\left(x(i)-x(i-1)\right)$. The  
projected dislocation density $\rho^{\rm p}$(i) is the projection  
of the density $\rho^{\rm II}$(i) from plane II onto plane I in order to deal 
with the non-parallel components
of the displacement. The  
$\gamma$-surface, $\gamma_3$, which in general includes shear-tension coupling
can be determined from {\it ab initio} calculations. $\tau^{\rm I(II)}_l$
is the external stress component interacting with the corresponding 
dislocation densities, 
$\rho^{\rm I(II)}_l(i)$ ($l$ = 1,2,3). This term represents 
the contribution to the total energy from the elastic
work done by the applied stress \cite{Bulatov,Lu1}. The response of 
a dislocation 
to an applied stress is achieved by the minimization of the energy 
functional with respect to
$\rho^{\rm I(II)}_l(i)$ at the given value of $\tau^{\rm I(II)}_l$. 
The dislocation core energy is defined as
the configuration-dependent part of the total energy, 
which includes the density-dependent part of the elastic energy and 
the entire misfit energy, in the absence of stress \cite{Lu1}.
$K_e$ and $K_s$ are
the edge and screw components of the general prelogarithmic elastic energy
factor $K$ \cite{Bulatov, Lu1}, while
 $\chi_{ij}$, $A_{ij}$, $B_{ij}$, $C_{ij}$ and $D_{ij}$ are 
double-integral kernels defined by
\begin{eqnarray*}
\chi_{ij}&=&\int\limits_{x_{j-1}}^{x_j}\int\limits_{x_{i-1}}^{x_i}
{\rm ln}|x-x'|dxdx',\\
A_{ij}&=&\int\limits_{x'_{j-1}}^{x'_j}\int\limits_{x_{i-1}}^{x_i}
\frac{1}{2}{\rm ln}(x_0^2+y_0^2)dxdx',\\
B_{ij}&=&\int\limits_{x'_{j-1}}^{x'_j}\int\limits_{x_{i-1}}^{x_i}
{\rm ln}\frac{x_0^2}{x_0^2+y_0^2}dxdx',\\
C_{ij}&=&\int\limits_{x'_{j-1}}^{x'_j}\int\limits_{x_{i-1}}^{x_i}
{\rm ln}\frac{y_0^2}{x_0^2+y_0^2}dxdx',\\
D_{ij}&=&\int\limits_{x'_{j-1}}^{x'_j}\int\limits_{x_{i-1}}^{x_i}
{\rm ln}\frac{x_0y_0}{x_0^2+y_0^2}dxdx',
\end{eqnarray*}
where $x_0 = L-x+x'\cos\theta$, and $y_0 = -x'\sin\theta$.
The equilibrium structure of the dislocation is determined by minimizing
the total dislocation energy functional energy with 
respect to the dislocation density.  

In order to compare and understand 
the different cross-slip behavior in Al and Ag, we 
have carried out {\it ab initio} calculations for their $\gamma$-surfaces. 
In both calculations, we used a supercell containing 
six layers of atoms in the [111] direction. The {\it ab initio} calculations 
are based on the pseudopotential plane-wave method \cite{Payne} 
within the local density approximation. 
Owing to the planar nature of the dislocation
core structure in fcc metals, we disregard in the $\gamma$-surface
calculations the displacement perpendicular 
to the slip planes and consider partially the shear-tension coupling by
performing volume relaxation along the [111] direction. 
The complete $\gamma$-surface
for Al and Ag is shown in Fig. 2(a) and 2(b)
respectively. The most striking difference between the two
$\gamma$-surfaces is the large difference in intrinsic stacking
fault energy, which is 165 mJ/m$^2$ for Al and 14 mJ/m$^2$ for Ag. 
This dramatic difference in $\gamma$-surface gives rise to very different 
dislocation core structures and cross-slip behavior that we are going to 
explore. The {\it ab initio} calculated Burgers vector, $b$, 
of a perfect 1/2[101] dislocation in Al and Ag is
2.85 and 2.83 \AA, respectively.

The model calculation is set up by introducing a screw dislocation at the 
intersection of the two slip planes without applying external stress 
to the system at first. The initial
configuration of the dislocation is specified by a step function for the screw 
displacement $f^{\rm I}_3(x)=0$ for $x<L$ and $f^{\rm I}_3(x)=b$ for $x \geq L$. 
All other displacement components including those on the cross-slip
plane are set initially to zero. This corresponds to a
pure screw dislocation with a zero width ``spread'' on the primary plane.
We then relax the dislocation structure according to the energy 
functional. The results of the dislocation density $\rho(x)$ 
in the primary and cross-slip planes
for Al and Ag are
presented in Figs. 3(a) and 3(b), respectively. 
The screw dislocation in Al which starts out at the 
primary plane spontaneously 
spreads into the cross-slip plane, as the density peak at the cross-slip
plane indicates. As expected, the edge component of the density is zero 
at the cross-slip plane because only the screw displacement can cross-slip.
On the other hand, the screw dislocation in Ag
dissociates into two partials, separated by 7.8 $b$ ($\approx$ 22 \AA), 
in excellent agreement with the experimental value of 
20 \AA~ in TEM measurements \cite{Cockayne}.
The left (right) partial has a positive (negative)
edge component of the Burgers vector represented by the positive (negative)
density. The integral of the edge density over all atomic
sites is zero, corresponding to a pure screw dislocation.    
These partial dislocations cannot cross-slip, as the arrows in Fig. 3(a) 
indicate, without first annihilating their edge components, and 
the dislocation density on the cross-slip plane is essentially zero. 
Apparently,
the lack of obvious dissociation in Al stems from the fact that Al
has a much higher intrinsic stacking fault energy than Ag. The absence 
of obvious dissociation into partials in Al is also consistent 
with experiment \cite{Duesb0}.

Next we apply an external Escaig stress to the dislocations and examine
the evolution of the dislocation core structure under stress with
the emphasis on the effect of stress on the dislocation cross-slip. 
The Escaig stress, defined as the edge component of the diagonal
stress tensor, interacts only with the edge displacement of a dislocation, 
extending or shrinking its stacking fault width depending on the sign
of the stress. We apply the Escaig stress only on the dislocation
at the primary plane, and the stress components projected on the
cross-slip plane are removed. The evolution of the dislocation core 
structure in Ag, 
represented by its displacement 
density distribution, is presented in Fig. 4 as the Escaig stress
is varied. Upon application of positive (stretching) stress of 
$\tau_1^{\rm I}$ = 0.32 GPa, the separation of the partials
rises rapidly from 7.8 $b$ (equilibrium separation in the absence 
of applied stress) to 12 $b$ (Fig. 4(a)).
In Fig. 4(a) we show only the density at the primary plane, as the
density at the cross-slip plane is essentially zero. For stress
higher than 0.32 GPa, the partials move to the
two ends of the simulation box and the lattice breaks down.  
To activate cross-slip, however, one needs to apply a negative 
(compressive) Escaig stress to the dislocation in order to annihilate 
the edge components of the partials' displacement, known 
as a constriction process. Upon application of negative stress, 
the partials move towards each other and reduce the width of the stacking
fault. During this process, the edge components of the displacement 
from the two partials annihilate each other at the primary plane 
while the screw component is being built up at both planes.
One example of such structure is shown in Fig. 4(b). The left and right
density peaks represent the original two partial dislocations, while
the third peak at the center corresponds to the build-up
of the screw density from the overlapping partials. 
Interestingly, the screw component of the dislocation density at the
cross-slip plane is also accumulating, indicating the inception
of the cross-slip process. However, further increasing the 
negative Escaig stress
does not yield smaller separation between the partials.  
We find that the lower limit for the partials separation in Ag 
is 1.7 $b$, which 
is in agreement with the atomistic simulation
result for Cu, reporting a corresponding value of 1.6 $b$ \cite{Duesbery}. 
In other words, no complete constriction can be achieved for a 
straight dislocation. The critical configuration in which the
constriction is most developed is shown in Fig. 4(c). In this case,
three partial dislocations with the same amount of screw component of the
Burgers vector are formed. The cross-slipped screw dislocation density
is about one third of the remaining screw density at the primary plane.
In order to complete the cross-slip, either thermal fluctuations or 
other type of external stress have to be present. This is because the
remaining edge component of the partials interacts with the Escaig
stress, and as a result the partials exchange signs and move away
from each other. For example,  
as shown in Fig. 4(d), the left partial now acquires a negative
edge density and the right partial a positive
edge density. Associated with the inversion 
of the edge component of the density for the partials,
a run-on stacking fault is formed between them,
with an energy of about 1.0 J/m$^2$. The run-on stacking fault is
the most unstable stacking fault in an fcc lattice, with atoms
from the neighboring (111) planes sitting right on top of each other.
The distance between the two partials is
more than 10 $b$. It is interesting to note that in the wake of the 
(partial) constriction process, a pure screw 
dislocation segment is formed at the intersection of the two 
planes, with an appreciable amount of cross-slip. 

Next we examine the situation of Al. We first apply
positive Escaig stress to the complete screw dislocation.
We find that the screw dislocation remains unsplit 
until the stress reaches the threshold value of 0.96 GPa, 
required to separate the 
overlapping partials. The dislocation 
core structure corresponding to such a stress-driven dissociation 
is shown in Fig. 5(a). We find that the screw density
component at the cross-slip plane is only reduced by a few
percent due to the small splitting, leaving the density
at the cross-slip plane approximately equal to that in the
primary plane. Increasing the positive stress will further
separate the partials and reduce cross-slip. The maximum
positive Escaig stress, however, that the dislocation can 
sustain is 1.92 GPa (Fig. 5(b)). In this configuration,
the partials are separated by 5 $b$. The central peak in the
screw density plot corresponding to the original complete dislocation
is reduced significantly, and the cross-slipped screw density
amounts to only 1/3 of the screw density at the primary plane.
Therefore, the application of positive Escaig stress in Al 
corresponds to an ``inverse cross-slip'' process that transfers
displacement from the ``cross-slip'' plane to the ``primary
plane''. Interestingly, applying negative Escaig stress to the
dislocation has no effect on the dislocation splitting or core
width. The dislocation remains unsplit all the way until
the stress is great enough to break down the lattice. Finally, 
it is important to point out that for all cases studied here 
in Al and Ag, upon removal of the applied external 
stress, the dislocation 
returns spontaneously to its equilibrium configuration.
            
We have also estimated the critical energetics that 
are relevant to cross-slip. For example, we calculated the 
constriction energy, defined as the difference in dislocation
core energy between the normal and constricted states. By approximating
the state with 1.7 $b$ separation between the partials as 
the constricted state,
we were able to estimate the
constriction energy for Ag to be 0.14 eV/$b$. 
A similar approach has
been used to evaluate the constriction energy for a screw dislocation in
Cu based on atomistic
simulations, reporting a value of
 0.17 eV/$b$, which is in a good agreement with our
model calculations \cite{Duesbery}. Obviously, the constriction 
energy for Al is zero because its normal state is fully constricted. 
We have also calculated the critical stress for cross-slip, which is defined 
as the glide stress in the cross-slip plane to move a partially constricted 
dislocation from the primary to the cross-slip plane 
\cite{Duesbery}. We find that the critical stress for 
cross-slip in Ag is 1.68 GPa, compared to 0.32 GPa
in Al. Finally, we estimated the cross-slip energy 
barrier, which in the context of our calculations, is defined as the
difference in dislocation core energy before and after cross-slip
takes place under the application of the above mentioned critical stress for
cross-slip. In other words, we calculate the core energy difference
for the dislocation between its normal state and the state that the dislocation
just starts to cross-slip under the critical cross-slip stress. 
We find that the cross-slip energy barrier in Ag is
0.14 eV/$b$, much larger than that of 0.05 eV/$b$ in Al. 
One needs to be cautious when comparing our 
results for the cross-slip energy barrier directly with experiment, 
since the dislocations are assumed to be straight in our current 
implementation of the Peierls-Nabarro model. However, it is possible to
extend the present formalism to deal with an arbitrarily curved dislocation
where a more realistic cross-slip energy barrier can be obtained. 
Nevertheless, the present model is still capable to provide reliable 
energetics for straight dislocations. 
 
In summary, we have presented a novel model 
based on the semidiscrete  
Peierls-Nabarro framework that allows the study of dislocation
cross-slip and constriction. The
$\gamma$-surface entering the model is determined from {\it ab initio}
calculations which provide reliable atomic interactions across the
slip plane.  We find that the screw dislocation in
Al can spontaneously spread into the cross-slip plane, while 
in Ag it dissociates into partials and can not cross-slip. 
We have also examined in detail the response of the dislocation core structure 
to an external Escaig stress  and the effect of negative Escaig 
stress on the constriction of the Shockley partials.
We find that one can not achieve 100\% constriction 
for the case of straight partial dislocations considered in this work.
By computing the dislocation core energy under stress we
estimate the dislocation constriction energy for Al and Ag. 
 The calculated values of the critical stress and the energy barrier 
for dislocation cross-slip demonstrate that dislocation 
cross-slip is much easier in Al than in Ag. Since our 
{\it ab initio} based model is much more expedient than 
direct {\it ab initio} atomistic simulations, it 
can serve as a powerful and efficient tool for alloy design, 
where the goal is to select the ``right'' elements with 
the ``right'' alloy composition to tailor desired 
mechanical, and in particular, dislocation properties, such as
cross-slip properties.  

\begin{acknowledgments}
Two of us (G.L. and N.K.) acknowledge the support from  
Grant No. DAAD19-00-1-0049 through the U.S. Army Research Office. G.L. 
was also supported by Grant No. F49620-99-1-0272 through the U.S. Air 
Force Office for Scientific Research.  
\end{acknowledgments}

\begin{figure}
\includegraphics[width=300pt]{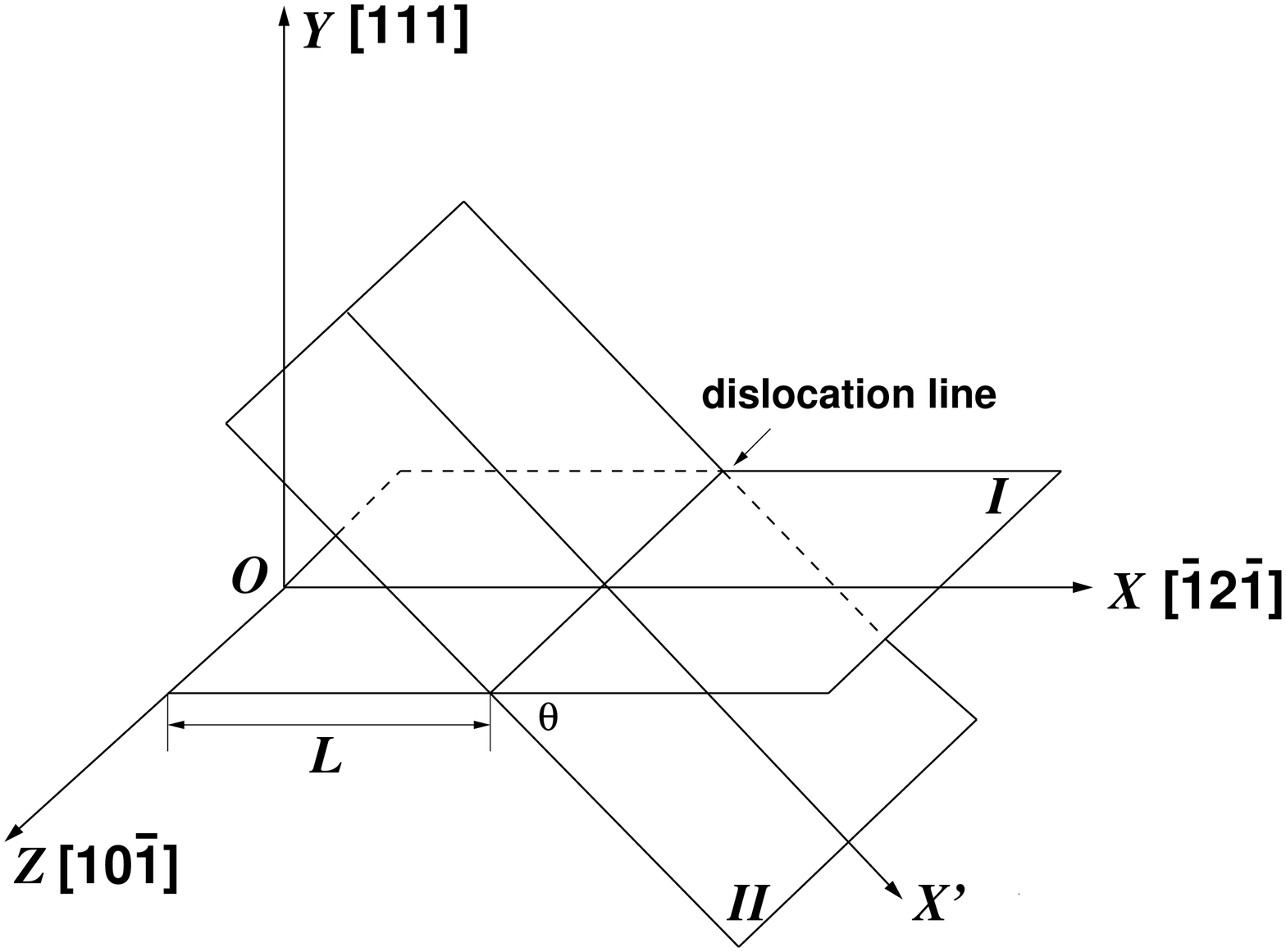}
\caption{Cartesian set of coordinates showing the directions relevant 
to the screw dislocation
located at the intersection of the two slip planes. Plane I (II) denotes the 
primary (cross-slip) plane.}
\label{fig1}
\end{figure}

\begin{figure}
\includegraphics[width=350pt]{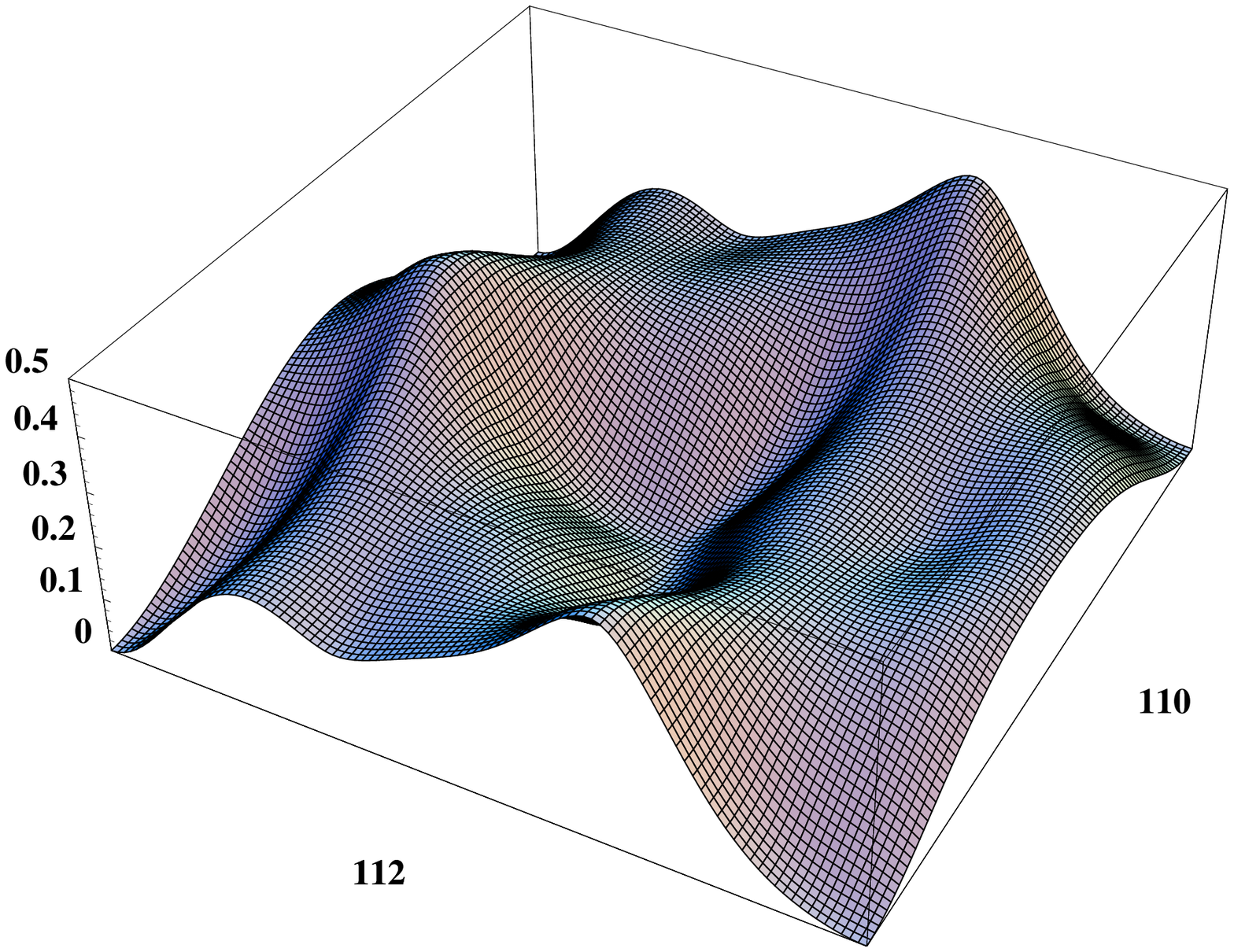}
\includegraphics[width=350pt]{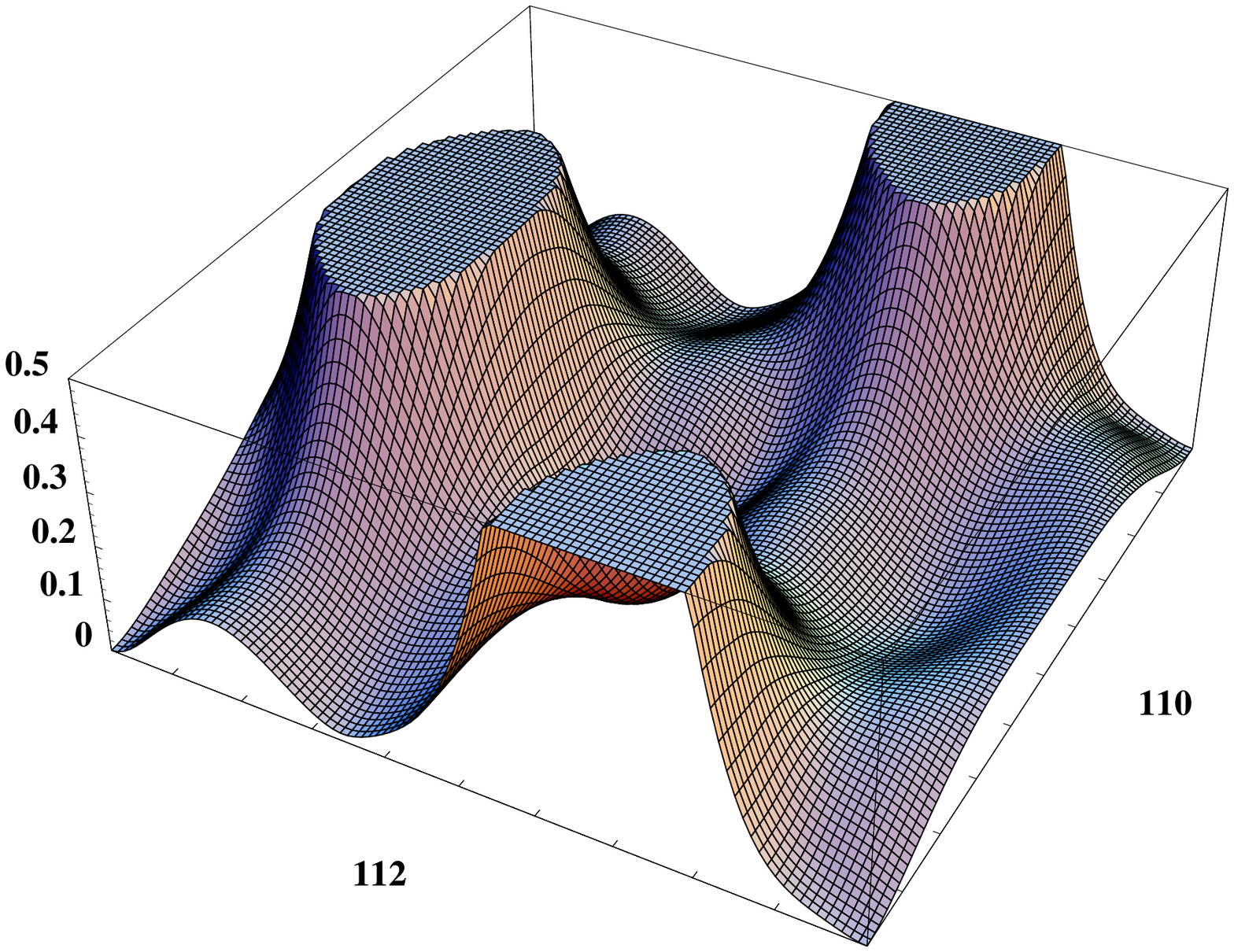}
\caption{The $\gamma$-surfaces (J/$m^2$) for displacements along a (111) plane
for (a) Al and (b) Ag .
The corners of the plane and its center correspond to
identical equilibrium configurations,
i.e., the ideal lattice.
The two energy surfaces are displayed
in exactly the same perspective and on the
same energy scale to facilitate comparison of important features.}
\label{fig2}
\end{figure}

\begin{figure}
\includegraphics[width=300pt]{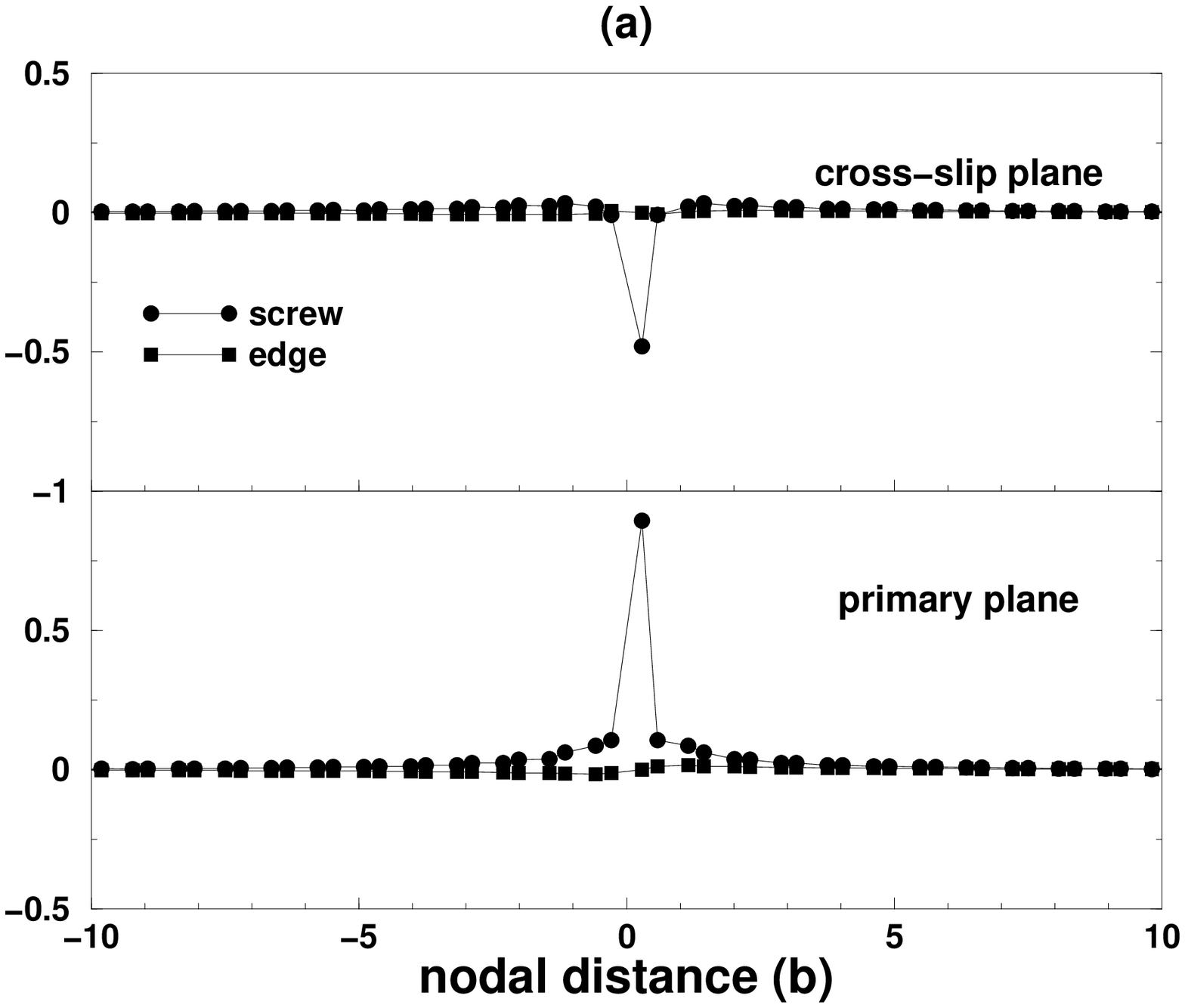}
\includegraphics[width=300pt]{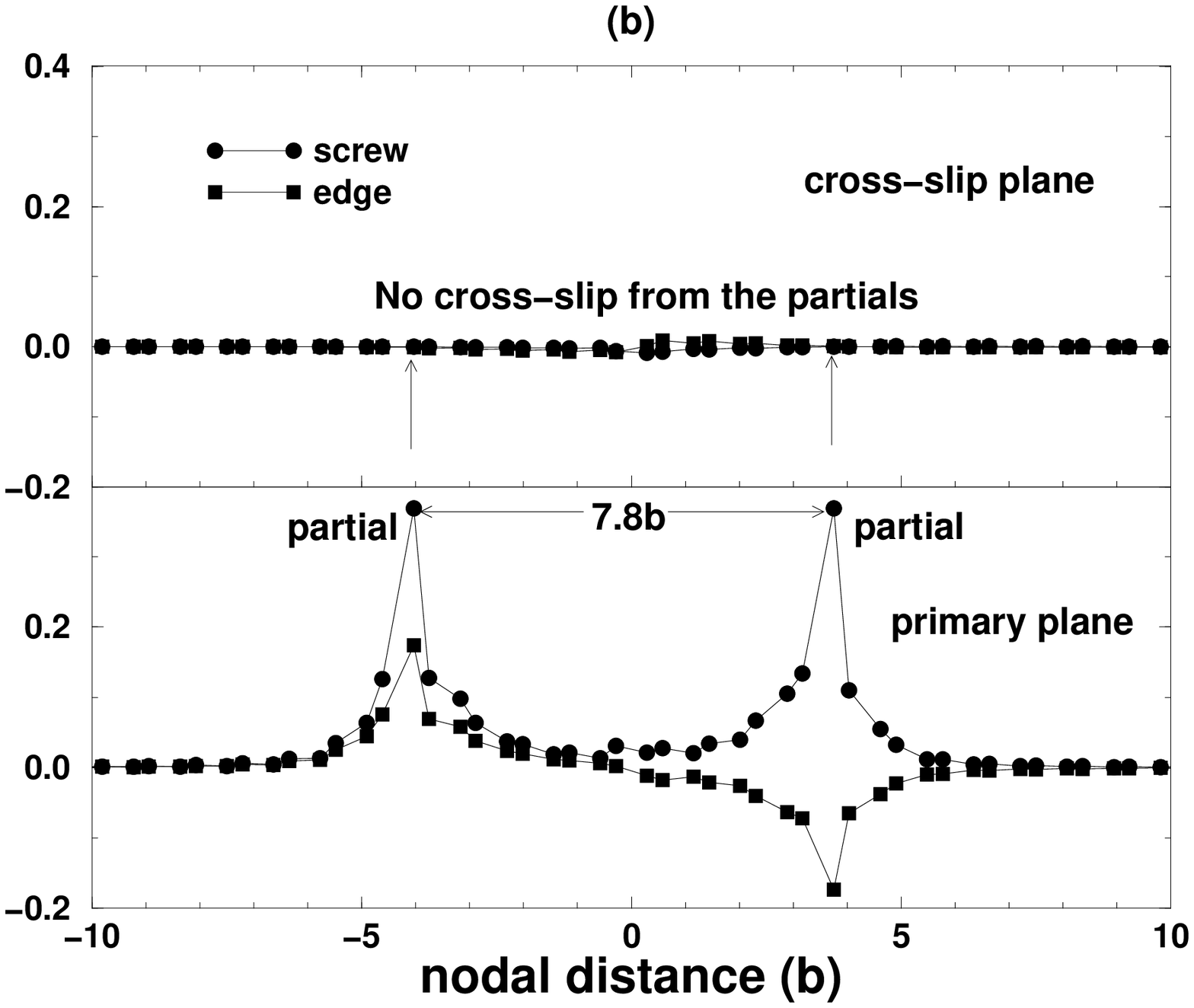}
\caption{Dislocation displacement density $\rho$(x) for Al (Fig. 3(a)) and Ag (Fig. 3(b)). 
The peaks in the density plot in Fig. 3(b) represent partial dislocations.}
\end{figure}

\begin{figure}
\includegraphics[width=400pt]{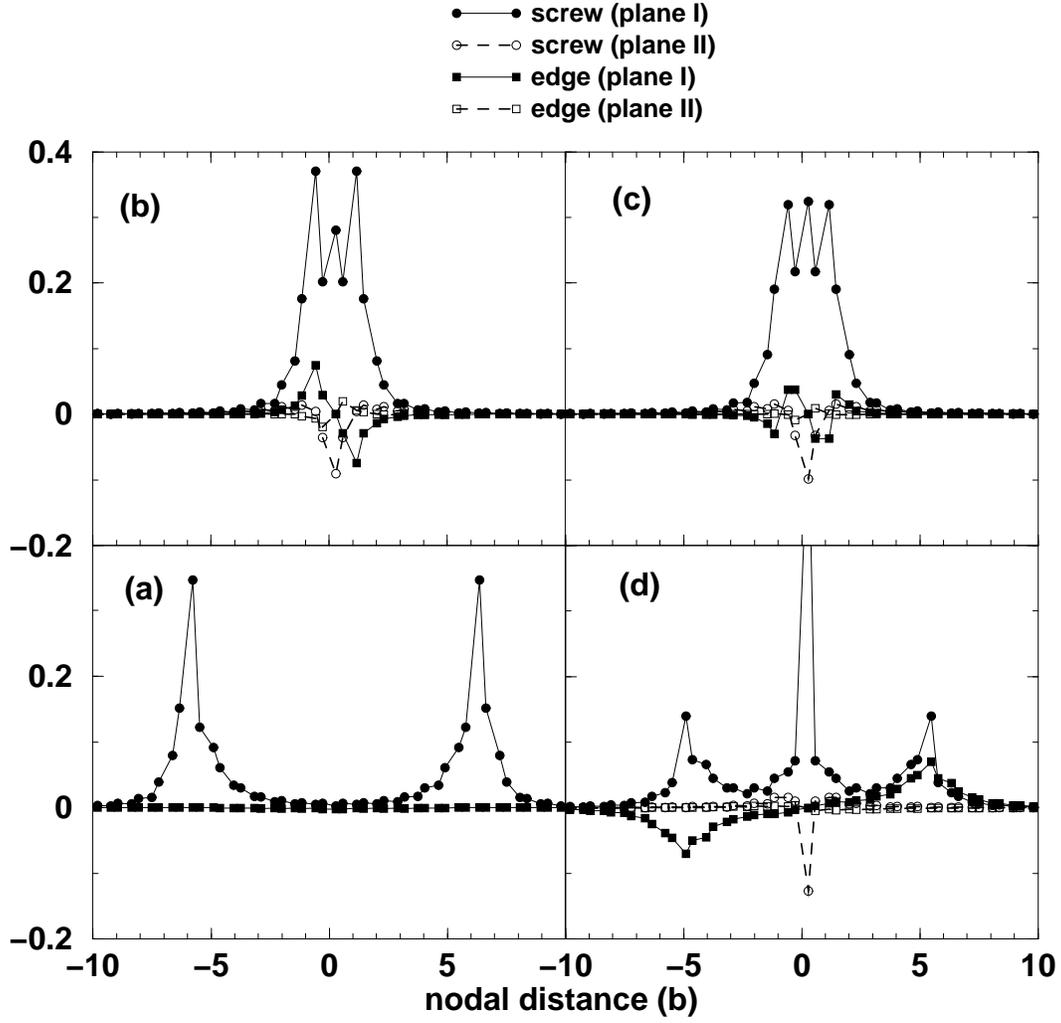}
\caption{Displacement density $\rho$(x) for the screw dislocation 
in Ag under different Escaig stress.
The solid (dashed) line and closed (open) symbols represent the dislocation density 
at the primary (cross-slip) plane. The screw and edge components of 
$\rho$(x) are represented by circle and square, respectively. 
The external Escaig stress is 
(a) $\tau_1^{\rm I}$ = 0.32 GPa, (b) $\tau_1^{\rm I}$ = -0.48 GPa,
(c) $\tau_1^{\rm I}$ = -0.64 GPa, and (d)  $\tau_1^{\rm I} $ = -3.2 GPa, 
respectively.  }
\end{figure}

\begin{figure}
\includegraphics[width=300pt]{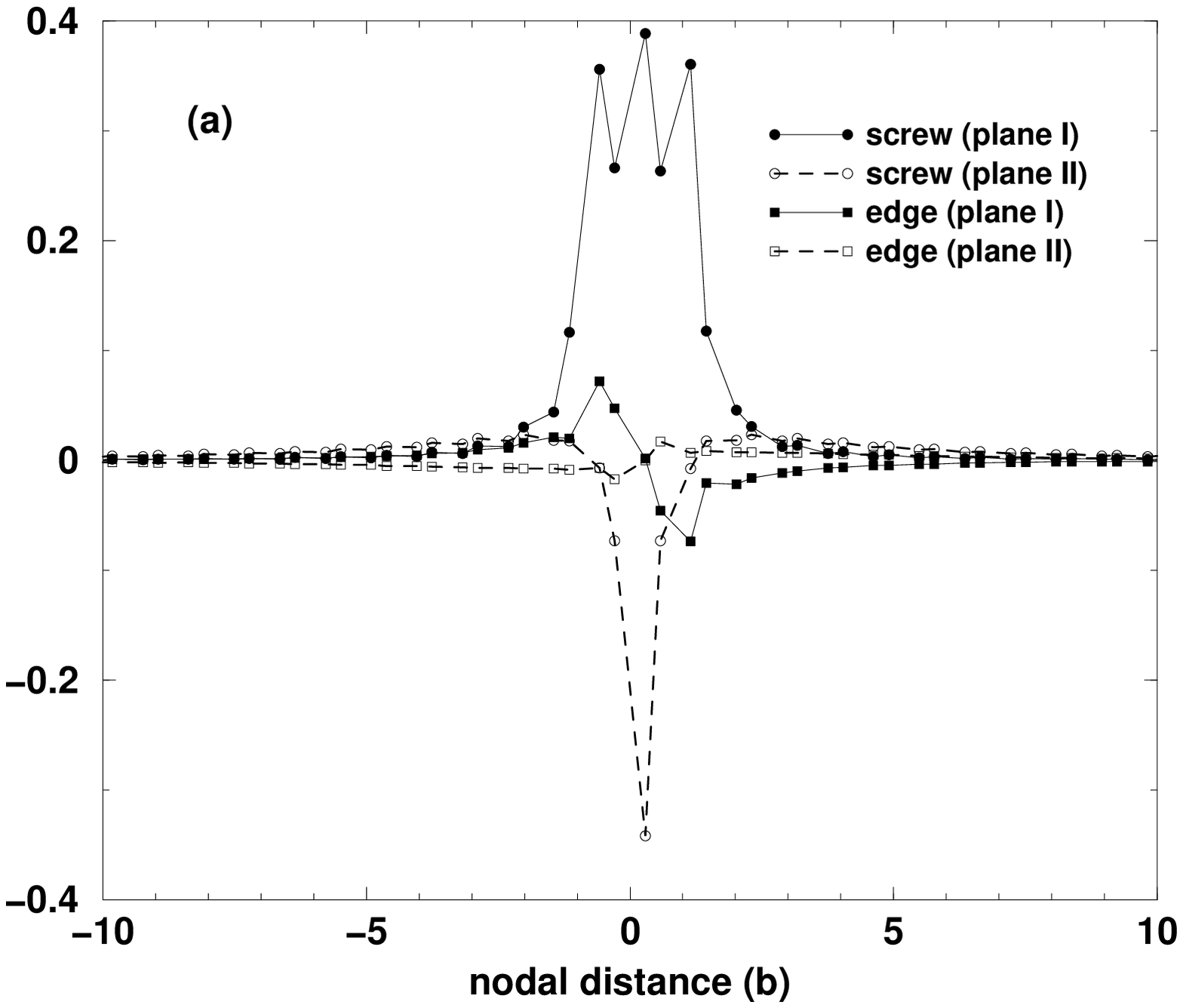}
\includegraphics[width=300pt]{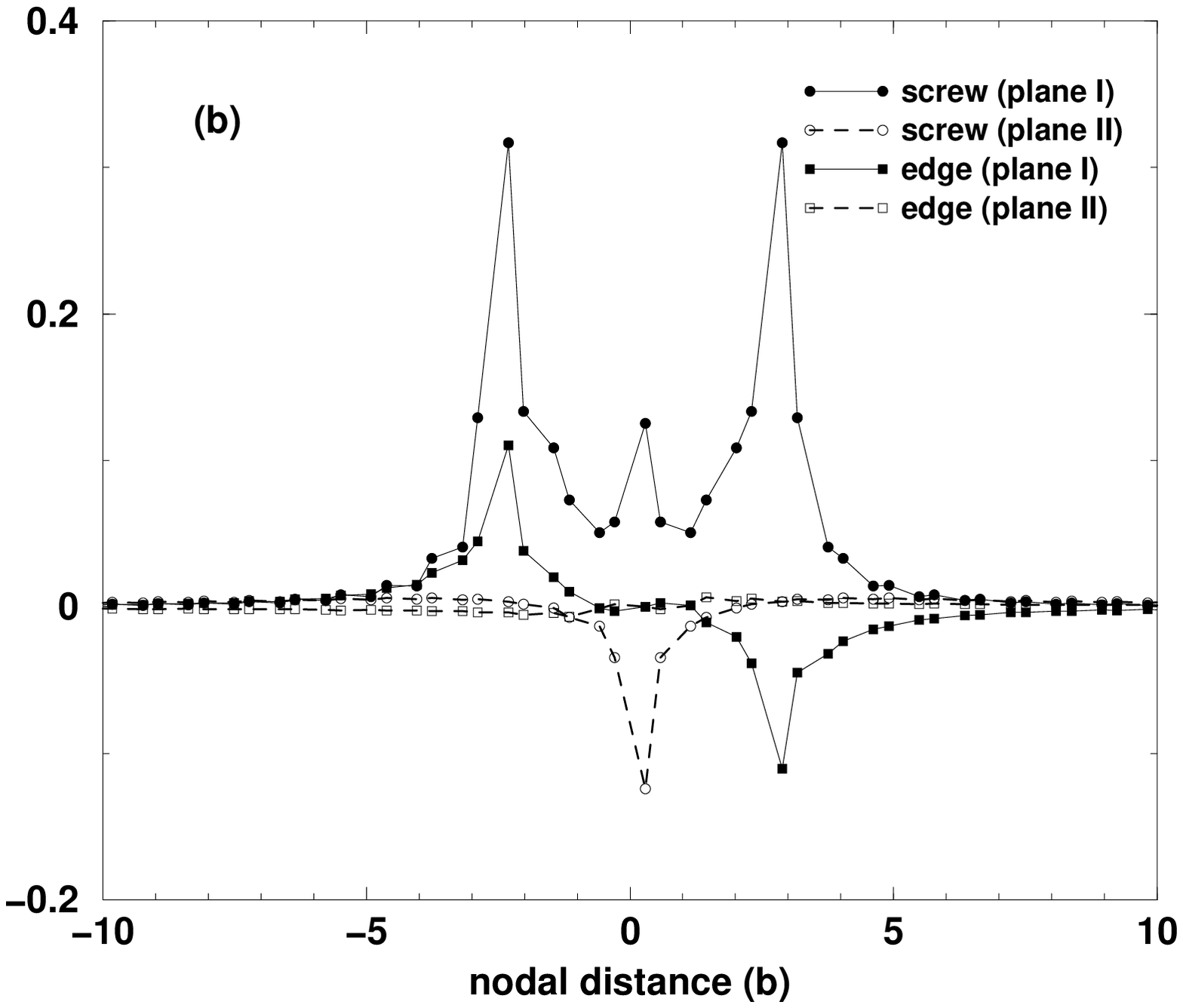}
\caption{Displacement density $\rho$(x) for the screw dislocation 
in Al under different Escaig stress.
The solid (dashed) line and closed (open) symbols represent the dislocation density 
at the primary (cross-slip) plane. 
The circle and square represent the screw and edge components
of $\rho$(x), respectively. The external Escaig stress is 
(a) $\tau_1^{\rm I} $ = 0.96 GPa and (b) $\tau_1^{\rm I} $ = 1.92 GPa, respectively.}
\end{figure}
\end{document}